\documentclass[11pt]{article}

\usepackage[matrix,arrow]{xy}

\usepackage{epsfig}

\usepackage{amssymb,amsthm}
\usepackage{amsmath}
\usepackage{amscd}
\usepackage{latexsym}
\usepackage{graphics}
\usepackage{color}

\usepackage{graphicx}

\input{xy}
\xyoption{all}

\catcode`@=11
\renewcommand{\section}
{\@startsection{section}{1}{0pt}{\medskipamount}{\medskipamount}{\large\bf}}
\makeatletter\renewcommand{\subsection}
{\@startsection{subsection}{2}{\z@}{-3.25ex plus -1ex minus -.2ex}
{1.5ex plus .2ex}{\it }}
\makeatletter\renewcommand{\subsubsection}
{\@startsection{subsubsection}{3}{\z@}{-3.25ex plus -1ex minus -.2ex}
{1.5ex plus .2ex}{\noindent\underline}}

\numberwithin{equation}{section}
\catcode`@=12

\def\={\ =\ }
\def\dd{{\rm d}}

\newcommand{\Tr}[1]{\:{\rm Tr}\,#1}
\def\e{{\,\rm e}\,}

\newcommand{\vech}[1]{\ensuremath{\boldsymbol{\hat{#1}}}}
\newcommand{\kb}[2]{|#1\rangle\langle#2|}
\newcommand{\ket}[1]{|#1\rangle}
\newcommand{\bra}[1]{\langle #1|}
\newcommand{\bk}[2]{\langle #1|#2\rangle}

\DeclareMathOperator{\End}{End}

\def\ii{{\,{\rm i}\,}}

\newcommand{\bbr}{\mathbb{R}}

\newcommand{\bbc}{\mathbb{C}}

\newcommand{\eps}{\mathcal{E}}

\def\hil{{\mathcal H}}
\def\bun{{\mathcal E}}

\def\e{\epsilon}

\def\beq{\begin{equation}}
\def\eeq{\end{equation}}
\def\bea{\begin{eqnarray}}
\def\eea{\end{eqnarray}}

\renewcommand{\e}{\,\mathrm{e}\,}

\newcommand{\R}{{\mathbb{R}}}
\newcommand{\N}{{\mathbb{N}}}
\newcommand{\C}{{\mathbb{C}}}

\def\End{{\rm End}}

\def\>{\rangle}
\def\<{\langle}
\def\+{\dagger}
\def\={\ =\ }

\begin{document}

\begin{titlepage}
\setcounter{page}{0}
\begin{flushright}
ITP--UH--03/10\\
HWM--10--1\\
EMPG--10--01\\
\end{flushright}

\vskip 1.8cm

\begin{center}

{\Large\bf UV/IR duality in noncommutative quantum field
  theory}\footnote{Based on invited talks given by the second author at the
  Bayrischzell Workshop on ``Noncommutativity and Physics: Quantum Geometries and Gravity'',
Bayrischzell, Germany, May 15--18 2009; at the 2nd School on ``Quantum Gravity and Quantum Geometry''
session of the 9th Hellenic School on Elementary Particle Physics and
Gravity, Corfu, Greece, September 13--20 2009; and at the Dublin
Institute for Advanced Studies Workshop on ``Noncommutativity and
Matrix Models'', Dublin, Ireland, November 23--27 2009. To be published
in {\sl General
Relativity and Gravitation}.}

\vspace{15mm}

{\large Andr\'e Fischer}\\[0.2cm]
 \noindent{\em Institut f\"ur Theoretische Physik\\
  Leibniz Universit\"at Hannover\\
  Appelstra\ss e 2, D-30167 Hannover,
    Germany}\\[+0.2cm] 
{Email: {\tt 
    afischer@itp.uni-hannover.de}}\\[1cm] 
{\large Richard J. Szabo}
\\[2mm]
\noindent {\em Department of Mathematics\\ Heriot-Watt University\\
Colin Maclaurin Building, Riccarton, Edinburgh EH14 4AS, U.K. \\ and Maxwell Institute
  for Mathematical Sciences, Edinburgh, U.K.}
\\[2mm]
{Email: {\tt R.J.Szabo@ma.hw.ac.uk}}

\vspace{30mm}

\begin{abstract}

\noindent
We review the construction of renormalizable noncommutative euclidean
$\phi^4$-theories based on the UV/IR duality covariant modification of
the standard field theory, and how the formalism can be extended to
scalar field theories defined on noncommutative Minkowski space.

\end{abstract}

\end{center}
\end{titlepage}



\section{Renormalization of noncommutative euclidean scalar field theory\label{RNCESFT}}

\noindent
In this article we will review recent progress in understanding how to
renormalize field theories on noncommutative euclidean space, and some
new advances into how these models may be analytically continued to
Minkowski signature. Here we will be exclusively interested in scalar
field theories on Moyal spacetimes of even dimension. If $\phi$ is a
real scalar field on $\R^{2d}$ with Fourier transform
$\widetilde\phi$, then the interactions in noncommutative field theory
on this space can be encoded by modifying the pointwise products
$\phi\cdot\phi$ to star-products $\phi\star\phi$, which in momentum
space amounts to altering the Fourier convolution products as
\beq
\widetilde\phi(k)\,\widetilde\phi(q)~\longrightarrow~
\widetilde\phi(k)\,\widetilde\phi(q)~
\e^{\ii k\times q} \ , \qquad k\times q\=\mbox{$\frac12$}\,
k_\mu\,\theta^{\mu\nu}\,q_\nu
\eeq
where $\theta^{\mu\nu}$ is a constant antisymmetric matrix which we
assume is of maximal rank for simplicity. Foundational aspects of the
theory are covered in~\cite{Szabo1}.
\unitlength=1.00mm
\linethickness{0.4pt}

In $\lambda\,\phi_{2d}^{\star n}$-theory, the interaction vertex in
momentum space is thus modified to
\beq
\begin{picture}(100.00,25.00)
\thinlines
\put(32.00,12.00){\line(-1,1){10.00}}
\put(32.00,12.00){\line(-1,-1){10.00}}
\put(32.00,12.00){\circle*{1.50}}
\put(18.00,22.00){\makebox(0,0)[l]{{$k_1$}}}
\put(18.00,2.00){\makebox(0,0)[l]{{$k_2$}}}
\put(26.00,12.00){\makebox(0,0)[l]{{$\lambda$}}}
\put(32.00,12.00){\line(1,1){10.00}}
\put(32.00,12.00){\line(1,-1){10.00}}
\put(40.00,12.00){\makebox(0,0)[l]{$\vdots$}}
\put(44.00,2.00){\makebox(0,0)[l]{{$k_3$}}}
\put(44.00,22.00){\makebox(0,0)[l]{{$k_n$}}}
\put(48.00,12.00){\makebox(0,0)[l]{{$~=~\lambda
\exp\Big(\ii\sum\limits_{I<J}\, k_I\times k_J\Big) \ . $}}}
\end{picture}
\label{intvertex}\eeq
The perturbative quantum field theory suffers from the infamous UV/IR
mixing problem~\cite{Minwalla99}. While the convergence properties of
planar graphs are the same as in the corresponding commutative quantum
field theory at $\theta=0$, the phase factors in (\ref{intvertex})
drastically alter the properties of non-planar graphs. Although this
phase factor improves the ultraviolet behaviour of amplitudes,
divergences reappear as poles at vanishing external momenta. This can
be summarized by saying that an ultraviolet cutoff {$\Lambda$}
necessarily induces an effective infrared cutoff
  {$\Lambda_0\=\frac1{\theta\,\Lambda}$}. This would seem to ruin
  standard renormalization schemes, such as the wilsonian prescription
  which requires a clear separation of energy scales. More precisely,
  although at one-loop order the on-shell amplitudes for massive
  particles are all finite, when these graphs are inserted as
  sub-graphs at higher-loop orders virtual particles of vanishing
  momentum produce uncontrollable divergences in the amplitudes. Thus the
  field theory cannot be renormalized. UV/IR mixing also occurs on
  more complicated noncommutative spaces such as $\kappa$-deformed
  spaces~\cite{Grosse06}, and thus appears to be a generic feature of
  noncommutative field theories.

The cure to this problem~\cite{Langmann02,Grosse04} is to consider
instead a covariant version of the field theory which renders its
ultraviolet and infrared regimes indistinguishable. In momentum space,
this can be regarded as a modification of the Fourier momenta as
\beq 
k_\mu~\longmapsto~K_\mu\=k_\mu+B_{\mu\nu}\,x^\nu \ ,
\label{Landaumom}\eeq
where $B_{\mu\nu}$ is another antisymmetric constant matrix of maximal
rank which is generically independent of $\theta^{\mu\nu}$. We may
think of this matrix as a ``magnetic'' background, so that the momenta
(\ref{Landaumom}) can be regarded as ``Landau'' momenta. If
$(k_\mu,x^\nu)$ are canonically conjugate variables, then these new momenta generate
a ``noncommutative momentum space'',
\beq
[K_\mu,K_\nu]\=2\ii B_{\mu\nu} \ ,
\eeq
the familiar feature of physical momenta for charged particles
propagating in a constant magnetic field. Thus the position and momentum spaces are formally identical, and there is no longer any distinction between what is meant by ultraviolet or infrared.

The {\it Grosse--Wulkenhaar model} is then the real euclidean scalar
   {$\lambda\,\phi_{2d}^{\star4}$}-theory in a background harmonic
   oscillator potential. Analogously to (\ref{Landaumom}), this amounts to replacing the Laplace operator giving the kinetic term in the scalar field action according to 
\beq
\partial_\mu^2~\longmapsto~\partial_\mu^2+\frac{\omega^2}2\,\widetilde
 x_\mu^{\,2} \ , \qquad \widetilde x_\mu\=2\theta_{\mu\nu}^{-1}\,x^\nu \ .
\eeq
The quantum field theory is then symmetric under Fourier transformation of fields, which amounts to exchanging momenta with positions as {$k_\mu~\leftrightarrow~\widetilde x_\mu$}.

The renormalization properties of this modified noncommutative field theory may then be summarized as follows~\cite{Grosse04}--\cite{Gurau09}. The covariant model is \emph{renormalizable to all orders in
{$\lambda$}}. A crucial ingredient of the original renormalizability proof~\cite{Grosse04} is the fact that the quantum field theory can regulated and described by a {matrix model}, with natural cutoff the matrix size
  {$N$}. At {$\omega\=1$} the field theory is completely invariant under the UV/IR duality transformation, without any rescalings of the parameters, whence this point in parameter space is called the self-dual point. At this point, the beta-functions in both couplings $\lambda$ and $\omega$ vanish to all orders of perturbation theory, and thus the renormalized coupling flows to a finite
  bare coupling. This is analogous to what happens in a conformally
  invariant quantum field theory. It implies, in particular, that the
  duality covariant noncommutative field theory contains no Landau
  ghost (or renormalons), contrary to the usual commutative
  $\phi_4^4$-theory, and unlike non-abelian gauge theories this
  elimination is achieved without asymptotic freedom (but instead with
  ``asymptotic safety''). For these reasons, a {non-perturbative} completion of the quantum field theory is believed possible. See~\cite{Rivasseau07} for further details.

\bigskip

\section{UV/IR duality on noncommutative euclidean space\label{UVIRDIES}}

\noindent
In this article we will demonstrate how to construct an analogous
duality covariant scalar field theory on noncommutative Minkowski
space. For later comparison, and because some of the duality proofs
transcribe immediately to lorentzian signature, we will first review
the proof of euclidean duality in some detail and the ensuing matrix model
representation, following~\cite{Langmann02,Langmann04}. The
Grosse--Wulkenhaar model has also been formulated on solvable
symmetric spaces in~\cite{Bieliavsky08}, where the UV/IR duality is
interpreted in terms of metaplectic representations of the Heisenberg group.

\subsection{Classical duality}

The original duality covariant model considers charged scalar fields {$\phi(x)$} on euclidean space {$\bbr^{2d}$} with action
\beq
S[\phi]\=\int\,\dd^{2d}x~\Big(\phi^\dagger\,\left(D_\mu^2+\mu^2\right)\,\phi
+g^2\,\phi^\dagger\star\phi\star\phi^\dagger\star\phi\Big) \ ,
\label{Sphi}\eeq
where
\beq
D_{\mu}\=\mbox{$\frac{1}{\sqrt{2}}$}\,\left(-\ii
\partial_{\mu}+B_{\mu\nu}\,x^\nu\right)
\eeq
is the gauge-covariant derivative in the magnetic background. All of our considerations also apply to the other inequivalent quartic interaction $\phi^\dag\star\phi^\dag\star\phi\star\phi$, but for simplicity we focus only on the one given in (\ref{Sphi}). This action is invariant under the {duality transformation} of order~two given by
\begin{eqnarray}
\phi({x})&\longrightarrow& \widehat{\phi}({x})\=\sqrt{|\det(B)|}~
\widetilde\phi(B\cdot{x}) \ , \nonumber \\[4pt]
\theta&\longrightarrow&
\widehat{\theta}\=-4B^{-1}\,\theta^{-1}\,B^{-1} \ , \nonumber \\[4pt]
g&\longrightarrow&\widehat g\=2^d\,\big|\det(B\,\theta)\big|^{-1/2}\,g \ .
\label{dualitytransf}\end{eqnarray}
The proof is an elementary, but somewhat lengthy, calculation~\cite{Langmann02}. The statement for the quadratic terms is a straightforward application of the Parseval identity, while for the quartic terms it follows from the integral kernel representation of the star-product (valid on the space of Schwartz fields $\phi$). With the normalizations above the {self-dual} point, where the action is completely invariant on parameter space, is given by {$\theta\=2B^{-1}$}.

\subsection{Quantum duality}

While the UV/IR duality is relatively straightforward at the classical level, it is somewhat more involved at the quantum level. The quantum field theory is defined by the generating functional of connected Green's functions
\beq
\mathcal{G}(J)\=-\log\frac{Z[J]}{Z[0]} \ ,
\eeq
where $Z[J]$ is the partition function for the field theory coupled to external sources $J$ and $J^\dag$ given by the functional integral
\beq
Z[J]\=\int\,\mathcal{D}\phi~\mathcal{D}\phi^\dagger~\exp\Big(-S[\phi]-
\int\,\dd^{2d}x~\big(\phi^\dag\,J+\phi\,J^\dag\big)\Big) \ .
\label{ZJ}\eeq
Correlation functions are obtained by taking suitable functional derivatives with respect to $J$ and $J^\dag$. By Parseval's identity, the extra source couplings in (\ref{ZJ}) are invariant under (\ref{dualitytransf}), and so is the functional integration measure. Thus the generating functional is {\it formally} invariant under the duality transformation of Schwartz functions {$\phi~\longmapsto~\widehat\phi$} on {$\bbr^{2d}$},
\beq
\mathcal{G}(J;B,g,{\theta})\= \mathcal{G}(\,\widehat J;B,\widehat
g,\widehat{\theta}\,) \ .
\eeq
However, to substantiate this claim we need to make sense of the functional integration in (\ref{ZJ}), which nessitates regulated perturbation theory. This requires a duality invariant regularization {$\mathcal{G}~\longrightarrow~\mathcal{G}_\Lambda$} which we now describe.

For simplicity, let us consider the case $d=1$. The crux of the proof is to expand the scalar fields not in the standard plane wave eigenstates of the momentum operator $-\ii\partial_\mu$ (which requires distinguishing between ultraviolet and infrared), but instead in the ``matrix basis'' {$f_{n,m}\in L^2(\bbr^2)$,
    $n,m\=0,1,\dots$} of {Landau wavefunctions}
\beq
\phi(x)\=\sum_{n,m=0}^\infty\,f_{n,m}(x)~\phi_{n,m} \ ,
\eeq
which are eigenstates of the Landau hamiltonian with
\beq
D_\mu^2f_{n,m}\=2B\,\big(n+\mbox{$\frac12$}\big)\,f_{n,m}~=:E_n\,f_{n,m} \ .
\eeq
Note that each Landau level is infinitely degenerate. The second quantum number features because the Landau wavefunctions are simultaneously eigenstates of the Landau hamiltonian with the reflected magnetic field,
\beq 
D_\mu^2\big|_{B\to-B}f_{n,m}\=E_m\,f_{n,m} \ ,
\eeq
such that the sum $D_\mu^2+D_\mu^2\big|_{B\to-B}\=-\partial_\mu^2+(B\cdot x)_\mu^2$ gives the background harmonic oscillator potential which figures in the Grosse--Wulkenhaar model.

For a suitable cut-off function {$F:[0,\infty)\to[0,\infty)$}, with
$F(0)\=1$ and $F(s)\=0$ for all $s>s_0$ and some finite $s_0$, we now replace the free
  propagator in the Landau basis,
\beq
C(n,m)\=\big(E_n+\mu^2\big)^{-1} \ ,
\eeq
with the regulated propagator
\beq
C_\Lambda(n,m)\=
\big(E_n+\mu^2\big)^{-1}\,F\big(\Lambda^{-2}\,(E_n+E_m)\big)
\label{CLambda}\eeq
where $\Lambda\in\R$ provides an ultraviolet cutoff in $(n,m)$-space and $C_\Lambda\to C$ as $\Lambda\to\infty$. The argument of the cut-off function is the eigenvalue of the differential operator $D_\mu^2+D_\mu^2\big|_{B\to-B}$. In particular, the cut-off on the operator $-\partial_\mu^2$ truncates all high-momentum modes while the cut-off on the operator $(B\cdot x)_\mu^2$ truncates the long-distance modes. 

Since $F(s)\=0$ for $s>s_0$ and some finite $s_0\in(0,\infty)$, for $\Lambda$ finite the propagators (\ref{CLambda}) are non-zero only for $E_n+E_m$ smaller than a uniform upper bound, which from the forms of the Landau eigenvalues can happen only for finitely many values of $n,m$.
With this regularization, every Feynman diagram is of the schematic form 
\beq
\sum_{n_1,m_1,\ldots,n_K,m_K\geq0}~\prod_{k=1}^K\,
C_\Lambda(n_k,m_k)~\times~({\rm vertices})
\label{finitesums}\eeq
where the vertex factors will in general be complicated combinatorial quantities, but their explicit form is immaterial for our argument. Since the propagators in (\ref{finitesums}) are non-zero for only finitely-many $n_k,m_k$, every Feynman diagram is given by a finite sum, i.e. all Feynman amplitudes converge. This completes the proof of quantum duality.

\subsection{Matrix model representation}

A beautiful feature of the covariant quantum field theory is that it can be mapped exactly onto a matrix model, a manipulation which has no counterpart in ordinary field theory. For simplicity, we consider the two-dimensional model at the self-dual point, $d\=1$ and  $\theta\=2B^{-1}>0$ (see~\cite{Langmann04} for the general case). The crucial feature is the projector property of the Landau wavefunctions with respect to the star-product,
\beq
f_{n,m}\star f_{n',m'}\=\sqrt{\frac B{4\pi}}~ \delta_{m,n'}~f_{n,m'} \ , 
\eeq
together with $f_{n,m}{}^*\=f_{m,n}$ and the normalization
\beq
\int\,\dd^2x~f_{n,m}(x) \=\sqrt{\frac{4\pi}B}~\delta_{n,m} \ .
\eeq

The action functional (\ref{Sphi}) can thereby be expressed in the form
\beq
S[\phi]\=\Tr\Big(\phi^\dag\,\mathcal{B}\,\phi+\mu^2\,\phi^\dag\,
\phi+\tilde g^2\,\big(\phi^\dag\,
\phi\big)^2\Big) \ ,
\eeq
where in this formula $\phi\=(\phi_{n,m})$ is an infinite matrix,
$\tilde g^2\=B\,g^2/4\pi$, and $\mathcal{B}_{n,m}\=2B\,\big(n+\frac12\big)~\delta_{n,m}$. The quantum field theory thus has a {$U(\infty)$} symmetry {$\phi~\longrightarrow~U^\dag\,\phi\,U$}, and is the {$N~\longrightarrow~\infty$} limit of the {$N\times N$} complex 
  matrix model in an external field whose partition function is given by
\beq
Z_N\=\int~\prod_{n,m=1}^N\,\dd\phi_{n,m}~\dd\phi^\dag_{n,m}~
\e^{-S[\phi]} \ .
\eeq
This is an integral over a finite-dimensional space, and thus gives a constructive non-perturbative definition of the quantum field theory. Various exact integrability properties of the model in this representation, which is related to the Kontsevich--Penner matrix model (a hermitean matrix model in an external field with logarithmic potential), are described in~\cite{Langmann04}.

\bigskip

\section{Analytic continuation to Minkowski signature}

\noindent
The continuation of the duality covariant field theory to noncommutative Minkowski space is naively obtained by Wick rotation {$x^0~\longrightarrow~\pm\ii t$ plus an additional change $B_{0i}~\longrightarrow~\pm\ii E_i$}, giving the dynamics in an electromagnetic background. While this is wrong for a number of reasons, we shall see that many of our results can be obtained in hindsight via a careful continuation of this sort. The reason that this naive approach is not expected to work is that the perturbative dynamics of (non-covariant) noncommutative field theory {\it cannot} be
  obtained simply by Wick rotation~\cite{Bahns02}--\cite{Rim02}. In contrast to the commutative case, time-ordering factors and the two-point function do not combine into Feynman
  propagators in non-planar graphs with a noncommuting time direction. Because of this complication, the Dyson and Feynman expansions are distinct, and the renormalization properties in the Dyson series are very
  different. By developing the complicated rules of time-ordered perturbation theory on noncommutative spacetime, one can restore unitarity and causality of the quantum field theory. In fact, it has been suggested that UV/IR mixing may be far less severe or even
  absent in this case~\cite{Bahns09}. In the following we shall summarize results from the analysis of~\cite{Fischer08} which are carried out by defining the quantum field theory using a functional integral framework, rather than time-ordered perturbation theory.

\subsection{Results}\label{results}

We will begin by stating the main results from~\cite{Fischer08}, before going into the technical details of their derivation, which requires various notions from functional analysis and the theory of generalized functions. Again we restrict to the case of $1+1$ dimensions, corresponding to a pure electric background, the general case being a straightforward combination with the earlier euclidean analysis~\cite{Fischer08}.

\subsubsection*{Matrix basis}

There is a dense domain of scalar fields {$\phi\in\Phi\subset L^2(\bbr^2)$}
  and ``electric Landau wavefunctions'' {$f_{n,m}^\pm\in\Phi', \
    n,m\=0,1,\dots$} such that 
\beq
\phi(x)\=\sum_{n,m=0}^\infty\,f^+_{n,m}(x)~\phi^-_{n,m} \= \sum_{n,m=0}^\infty\,
f^-_{n,m}(x)~\phi^+_{n,m} \ ,
\label{phix}\eeq
and
\beq
D_\mu^2f^\pm_{n,m}\=\pm\ii E_n\,f_{n,m}^\pm
\ , \qquad D_\mu^2\big|_{B\to-B}f^\pm_{n,m}\=\pm\ii E_m\,f^\pm_{n,m} \ .
\label{Dfpmnm}\eeq
The $\pm$ labels here correspond to the two choices of sign in the Wick rotation. In fact, although we shall not derive them in this way, the wavefunctions $f_{n,m}^\pm$ can be obtained merely by Wick rotating the standard Landau wavefunctions. Nevertheless, this result should look somewhat odd to the reader, since (\ref{Dfpmnm}) seems to assert that $f_{n,m}^\pm$ are eigenfunctions of self-adjoint operators with \emph{imaginary} eigenvalues. However, the crucial point is that these functions live in the topological dual $\Phi'$ of the domain $\Phi$, which is much larger than the domain of these differential operators. Below we will see how this can be used to define notions of generalized eigenfunctions with generalized eigenvalues, which can be complex. The electric Landau wavefunctions obey the $L^2$-orthonormality and star-product projector relations
\beq
f_{n,m}^\pm\,^*\=f_{m,n}^\mp \ , \quad \big(f_{n,m}^\pm\,,\,
f_{n',m'}^\mp\big)_{L^2}\=\delta_{m,n'}\,\delta_{n,m'} \ , \quad
f^\pm_{n,m}\star f^\pm_{n',m'}\=\sqrt{\frac B{4\pi}} ~\delta_{m,n'}\,f^\pm_{n,m'} \ ,
\eeq
together with the normalization condition
\beq
\int\,\dd^2x~f_{n,m}^\pm(x)\= \sqrt{\frac{4\pi}B}~\delta_{n,m} \ .
\label{fpmnorm}\eeq
A better understanding of the physical meaning of these functions, and
how the formulation of the field theory in terms of them is related to
time-ordered perturbation theory, is currently lacking.

\subsubsection*{Unitarity and causality}

Each set of functions $f_{n,m}^+$ and $f_{n,m}^-$ on its own generates
a complete basis for expansion of Schwartz fields $\phi$. The domain
$\Phi$ is chosen such that both expansions can be taken
simultaneously. Both matrix bases together imply stability and CT-invariance during matrix regularization.
By stability we mean that expanding $\phi$
in the matrix bases $f_{n,m}^\pm$ and imposing the matrix regularization by cutting off these sums at some finite $N$ yields action functionals $S_\Lambda^\pm[\phi]$,
whose sum $S_\Lambda[\phi]=\frac12\, \big(S_\Lambda^+[\phi]+S_\Lambda^-[\phi]\big)$ is
manifestly real. The CT-symmetry $\phi_{n,m}^\mp\=C\,T\phi_{n,m}^\pm$ follows from
the behaviours of the electric Landau wavefunctions under $T$, $P\,T$
and $C$ transformations given by
\begin{eqnarray}
f_{m,n}^\pm(-t,x)&=&(-1)^{m-n}\,f_{n,m}^\pm(t,x) \ , \nonumber \\[4pt]
f_{m,n}^\pm(-t,-x)&=& (-1)^{m-n}\, f_{m,n}^\pm(t,x) \ , \nonumber \\[4pt]
f_{m,n}^\pm(t,x)^*&=&f_{n,m}^\mp(t,x)\=(-1)^{m-n}\, f_{m,n}^\mp(-t,x)\,.
\end{eqnarray}

\subsubsection*{Quantum duality}

The proof of classical duality follows exactly the same route as in
the euclidean case -- it does not depend on the signature of the inner
products used in Fourier transformation. At the quantum level, 
analogously to the euclidean case the regulated propagators in Minkowski
  space are obtained by replacing
\beq
C^\pm(n,m)\=\big\langle\,{\phi_{m,n}^\pm}^*\,
\phi_{m,n}^\mp\big\rangle
\label{Cpmnm}\eeq
with
\beq
C_\Lambda^\pm(n,m)\=
2\ii\,\big(\pm\ii E_n+\mu^2\big)^{-1}\,F\left(\Lambda^{-2}\,
|E_n+E_m|\right) \ .
\eeq
Then the proof of quantum duality presented before carries through verbatim
using these two sets of two-point functions. By
multiplying (\ref{Cpmnm}) with $f_{m,n}^\pm(x)^*\,f_{m,n}^\mp(y)$ and summing over all
$n,m\in\N_0$ one obtains the
position space representation of the propagator~\cite{Fischer08}. For
the free Klein-Gordon field without electric field, after Fourier
transformation this representation can be shown to possess the standard physical
mass-shell poles~\cite{Fischer10}.

\subsubsection*{Coupled complex two-matrix model representation} 

A non-trivial interacting two-matrix model now describes the
minkowskian theory, whose action is generically rather
involved~\cite{Fischer08}. With the same notation as before, it simplifies at the self-dual point to
\beq
S[\phi]\=\frac12\,\sum_{\pm}\,\Tr\Big(\pm\,\phi_\pm^\dag\ii\mathcal{B}\,
\phi_{\mp}+\mu^2\,\phi_\pm^\dag\,\phi_{\mp}+\tilde g^2\,\big(
\phi_\pm^\dag\,\phi_{\mp}\big)^2\Big) \ .
\label{Sphi2matrix}\eeq
This action possesses a much larger
{$GL(\infty)\times GL(\infty)$} symmetry
{$\phi_\pm~\longmapsto~\phi_\pm\,U_\pm, \
\phi_\pm^\dag~\longmapsto~U_{\mp}^{-1}\,\phi_\pm^\dag$}, and also the
discrete CT-symmetry
\beq
\big(\phi_\pm\,,\,\phi_\pm^\dag\,\big)~\longmapsto~
\big(\phi_{\mp}\,,\,\phi_{\mp}^\dag\big) \quad , \quad
\theta~\longmapsto~-\theta \ .
\eeq
This two-matrix model representation clearly demonstrates that the
Minkowski theory is not simply a Wick rotation of the euclidean theory.

\subsection{Derivations\label{deriv}}

We will now sketch how these results are obtained, including a description of the configuration space $\Phi$. Just like the analysis
of the standard Landau problem, and hence the duality covariant field theory
in euclidean signature, is related to the harmonic oscillator, the
model in Minkowski signature is related to the \emph{inverted
harmonic oscillator} whose hamiltonian is given by
\beq
H\=\mbox{$\frac12$}\,\big(p^2-\omega^2\,q^2\big) \ ,
\label{invham}\eeq
where $\omega\in\R$ and $(p,q)$ are canonically conjugate variables. The functional analytic properties of the corresponding
quantum hamiltonian are described in~\cite{Chruscinski03}. It is related to
the usual harmonic oscillator hamiltonian by a \emph{complex scaling}, which is a
non-unitary similarity transformation which sends
{$\omega~\longrightarrow~\pm\ii\omega$}. The first quantized operator
{$\hat H$} corresponding to (\ref{invham}) is
symmetric on a suitable domain in {$L^2(\bbr)$} with
spectrum {${\rm Spec}(\hat H)\=\bbr$. In fact it 
  admits a one-parameter family of self-adjoint extensions. This parameter should have some significance within the
  context of our duality covariant quantum field theory, but for
  simplicity we fix a self-adjoint extension and simply work with
  that. The relation between the classical hamiltonian $H$, the
  quantum hamiltonian $\hat{H}$, and the differential operator $D_\mu^2$ is established via the star
product and the Wigner transform, defined for a rank one operator
$\vech{\phi}=\kb{\psi}{\varphi}\in L^2(\R)\otimes L^2(\R)^\vee$ by
\beq
{\sf W}(\vech{\phi}\,)(t,x) \=\frac1{2\pi}\,
\int_{\R}\, \dd k~ \e^{\ii k\,x}\,\bk{t-\theta
   \,k/2}{\psi}\,\bk{\varphi}{t+\theta\, k/2} \ .
\label{Wigenfn}\eeq
For every function $f(x)\={\sf W}(\hat{f}\,)(x)$
one has
\beq
D_\mu^2 f(x)\=H\star f(x)\={\sf W}(\hat{H}\,\hat{f}\,)(x) \ .
\eeq
Thus instead of working with
$D_\mu^2$ operating on a set of fields, we will work with $\hat{H}$ acting on a
suitable quantum mechanical Hilbert space.

The hamiltonian $\hat H$ also has a set of {generalized eigenfunctions} with {\it imaginary}
eigenvalues. They occur as residues of the original eigenfunctions analytically
  continued to the complex energy plane. By closing the contour of
  integration in the eigenfunction expansion of a wavefunction appropriately, we pick up
  these states and obtain the analog of the {\it
    discrete} expansion in Landau wavefunctions. These techniques are
  analogous to those of the Bohm--Gadella theory of resonant
  states in quantum mechanics, wherein the instabilities
  mentioned above describe nuclear decay phenomena. The inverted
  harmonic oscillator potential and its resonance expansion also
  defines the analytic continuation of the Grosse--Wulkenhaar model to Minkowski
  signature. To explore the renormalization in this case, one needs to
  establish suitable decay properties of the free
  propagators in the matrix basis analogous to the euclidean
  case~\cite{Grosse04}. The properties of these Green's functions are
  currently under investigation~\cite{Fischer10}.

Let us now explain the concepts introduced above. The mathematical setting we need
is the extension of the notion of Hilbert space to that of a \emph{rigged Hilbert space} (also known in the literature
as a Gel'fand triple), which is a triple of spaces
\beq
\Phi~\subset~\hil~\subset~\Phi'
\eeq
where $\Phi$ is a dense nuclear subspace of a Hilbert space {$\hil$} with dual
{$\Phi'$}, the space of continuous linear functionals $\Phi\to\C$. If
$A\in\End(\hil)$ is self-adjoint on $\Phi$, then we can define its
action on $\Phi'$ using the dual pairing. A vector $F_\lambda\in\Phi'$
is then said to be a generalized eigenvector of $A$ with generalized
eigenvalue $\lambda\in\bbc$ if
\beq
\langle AF_\lambda|\phi\rangle~:=~ \langle
   F_\lambda |A\phi\rangle\= \lambda\,\langle F_\lambda|\phi\rangle
\eeq
holds for all $\phi\in\Phi$. The Gel'fand--Maurin theorem asserts
  that for any
  {$\phi\in\Phi$}, there exists
  {$F_\lambda\in\Phi'$} such that there is an expansion
\beq
\phi\=\int_{{\rm Spec}(A)}\,
    \dd\mu(\lambda)~F_\lambda\,\langle F_\lambda|\phi\rangle \
    ,
\eeq
where $\dd\mu$ is discrete measure on the discrete part of the
spectrum of $A$ and Lebesgue measure on the continuous part. 

For the example of the inverted harmonic
oscillator, the rigged Hilbert space is
\beq
\mathcal{S}(\bbr)~\subset~L^2(\bbr)~\subset
    ~\mathcal{S}'(\bbr) \ ,
\eeq
where $\mathcal{S}(\bbr)$ is the
  topological vector space of Schwartz functions on $\R$ (with the
  usual semi-norm topology) and $\mathcal{S}'(\bbr)$ is the space of
  tempered distributions on $\R$.
By parity invariance, each eigenvalue {$\bun\in{\rm Spec}(\hat H)=\R$}
corresponds to 
two-fold degenerate eigenfunctions {$\chi_\pm^\bun \ , \
    \eta_\pm^\bun\in \mathcal{S}'(\bbr)$} which after rescaling $B\to
  B/2$ are given explicitly by
\bea
\chi_\pm^\bun(q)&=&\frac{C}{\sqrt{2\pi \,B}}\, \ii^{\frac{\nu}{2}+\frac{1}{4}}\, \Gamma(\nu+1)\, D_{-\nu-1}\big(\mp\,\sqrt{-2\ii B}\,q\big) \ , \nonumber\\[4pt]
\eta_\pm^\bun(q)&=&\frac{C}{\sqrt{2\pi \,B}}\, \ii^{\frac{\nu}{2}+\frac{1}{4}}\, \Gamma(-\nu)\,D_\nu\big(\mp\,\sqrt{2\ii B}\,q\big) \ ,
\eea
where $C$ is a numerical constant, $\nu=-\ii\frac{\bun}{B}-\frac12$,
and $D_\nu(z)$ are parabolic cylinder functions. Only two 
of them are linearly independent, so for any {$\phi\in\mathcal{S}(\bbr)$}
the Gel'fand--Maurin theorem gives a pair of expansions
\beq
\phi(q)\=\sum_{\pm}~\int_\R\,\dd\bun~\chi^\bun_\pm(q)~\big\langle
\chi^\bun_\pm\,\big|\,\phi\big\rangle \=
\sum_{\pm}~\int_\R\,\dd\bun~\eta^\bun_\pm(q)~\big\langle 
\eta^\bun_\pm\,\big|\,\phi\big\rangle \ .
\label{phichieta}\eeq
The oscillator hamiltonian {$\hat H$} also has generalized eigenfunctions {$f_n^\pm$}
  with discrete eigenvalues
  {$\pm\,\ii B\,\big(n+\frac12\big) \ , \ n\=0,1,\dots$},
  occuring as residues of {$\chi_\pm^\bun \ / \ \eta_\pm^\bun$} 
  in the lower / upper complex half-plane. Then in a suitable domain
  {$\phi\in\Phi\subset\mathcal{S}(\bbr)$}, an application of the
  residue theorem to the energy integrals in (\ref{phichieta}) gives
  the respective \emph{resonance expansions}
\beq
\phi(q)\= \sum_{n=0}^\infty\,f_n^-(q)~
\big\langle f_n^+\big|\phi\big\rangle \= \sum_{n=0}^\infty\,f_n^+(q)~
\big\langle f_n^-\big|\phi\big\rangle \ .
\label{phiq}\eeq

The choice of configuration space $\Phi$ must ensure that both integrals over $\bun$
in (\ref{phichieta}) can be extended to closed contour integrals for
which the residue theorem applies in the usual way and such that the
resonance expansions in (\ref{phiq}) converge. In particular, it consists
of fields $\phi$ such that $\chi^\bun_\pm(q)\,\big\langle
\chi^\bun_\pm\,\big|\,\phi\big\rangle$ vanishes uniformly almost everywhere as $\bun$ tends to
infinity in the lower complex half-plane, and $\eta^\bun_\pm(q)\,\big\langle
\eta^\bun_\pm\,\big|\,\phi\big\rangle$ vanishes uniformly almost everywhere as $\bun$ tends to
infinity in the upper complex half-plane, together with the analogous
vanishing requirements on the scalar products
$\big\langle\psi\,\big|\,\chi^\bun_\pm\,\big\rangle \,\big\langle
\chi^\bun_\pm\,\big|\,\phi\big\rangle$ and $\big\langle\psi\,\big|\,
\eta^\bun_\pm\big\rangle \,\big\langle
\eta^\bun_\pm\,\big|\,\phi\big\rangle$ for all $\phi,\psi\in\Phi$. For this, consider the rigged Hilbert space
\beq
\mathcal{S}^\alpha_\alpha(\bbr)~\subset~L^2(\bbr)~\subset~
\mathcal{S}^\alpha_\alpha(\bbr)' \ ,
\eeq
where {$\mathcal{S}^\alpha_\alpha(\bbr)$} is a Gel'fand--Shilov space with
{$\alpha\geq\frac12$}, and its dual
{$\mathcal{S}^\alpha_\alpha(\bbr)'$} is a space of tempered
  ultra-distributions of Roumieu type. These Gel'fand--Shilov spaces
  contain entire functions ${\phi(q)}$
on {$\bbc$} restricted to {$\bbr$}, with $L^\infty$-norms obeying
{$\big\|q^m\,\partial_q^n\phi\big\|_\infty\leq
  C\,M^{n+m}\,n^{\alpha\,n}\,m^{\alpha\,m}$} for all $n,m\in\N_0$ with
some constant $C$ and given $M$. They form dense subspaces of Schwartz
space {$\mathcal{S}^\alpha_\alpha(\bbr)\subset \mathcal{S}(\bbr)\=
  \mathcal{S}^\infty_\infty(\bbr)$} which are closed under Fourier
transformation and the star-product~\cite{Soloviev07,Chaichian07}, and
which are generated by the basis of harmonic oscillator
wavefunctions~\cite{Lozanov07}. They are thus natural configuration
spaces for duality covariant noncommutative field theories. The
boundedness properties of these functions, together with the
asymptotic behaviours of the parabolic cylinder functions and the
gamma-functions, appear to be sufficient to ensure that the integrands in (\ref{phichieta})
and all pertinent pairing factors vanish appropriately~\cite{Fischer08,Fischer10}.

The mapping from functions on $\R$ in (\ref{phiq}) to fields on $\R^2$
in (\ref{phix}) is given by applying the Wigner transformation ${\sf
  W}\,:\,\mathcal{S}_\alpha^\alpha(\R)\otimes\mathcal{S}_\alpha^\alpha(\R)^\vee~\longrightarrow~ 
\mathcal{S}_\alpha^\alpha(\R^2)$ to a rank one operator
$\vech{\phi}=\kb{\psi}{\varphi}$ using the integral formula (\ref{Wigenfn}).
Expanding $\vech{\phi}$ in parabolic cylinder functions,
\begin{eqnarray}
 \vech{\phi} &=& \sum_{s,s'=\pm}~\int_{\R}\,
 \dd\eps~\int_{\R}\,\dd\eps'~\ket{\chi_s^\eps}\,\bk{\chi_s^\eps}{\psi}\,
\bk{\varphi}{\eta_{s'}^{\eps'}}\,\bra{\eta_{s'}^{\eps'}}
\nonumber\\[4pt] &=& \sum_{s,s'=\pm}~\int_{\R}\,
 \dd\eps~\int_{\R}\,\dd\eps'~\ket{\eta_s^\eps}\,\bk{\eta_s^\eps}{\psi}\,
\bk{\varphi}{\chi_{s'}^{\eps'}}\,\bra{\chi_{s'}^{\eps'}} \ , \label{chichi}
\end{eqnarray}
one can read off the respective resonance expansions (\ref{phix}) from
(\ref{phichieta}) and (\ref{phiq}). Using this mapping one can also explicitly compute the electric Landau wavefunctions~\cite{Fischer08}
\begin{eqnarray}
f_{m,n}^\pm(t,x)&=&
(-1)^{\min(m,n)}~\sqrt{\frac{B}{\pi}}~\sqrt{\frac{\min(m,n)!}{\max(m,n)!}}~(\pm\ii
B)^{|m-n|/2} \nonumber\\ && \times ~ \e^{\mp\ii
  B\,x_+\,x_-/2}\,x_{\mp\,{\rm sgn}(m-n)}^{|m-n|}\,
L_{\min(m,n)}^{|m-n|}(\pm\ii B\,x_+\,x_-) \ , \label{fmn}
\end{eqnarray}
where $x_\pm=t\pm x$ and
\beq
L_n^\alpha(z)\= \sum_{q=0}^n\, {\alpha+n\choose n-q} \, \frac{(-z)^q}{q!}
\eeq
are the generalized Laguerre polynomials. Using these explicit forms
one can straightforwardly derive all properties of $f_{m,n}^\pm$
stated above.

Essential for the proof of orthogonality is the
occurance of the phase factors $\e^{\mp\ii B\,x_+\, x_-/2}$ in (\ref{fmn}), which
generate derivatives via the identity $\int_\R\,\dd x_-
~(x_-)^p~\e^{\mp \ii x_+\, x_-} \= 2\pi\,(\pm\ii\partial_{+})^p\delta(x_+)$ and ensure that the integrals
over $(x_+,x_-)\in\R^2$ converge. This is the reason why only the
$L^2$-inner products $\big({f_{m,n}^\mp}\,,\,{f_{k,l}^\pm}\big)_{L^2}$ are permitted,
since this exponential factor is absent for the other
combinations. For the same reason only terms with equal powers of
$x_+$ and $x_-$ survive the integration. After some
algebra one then readily verifies the
orthogonality relations
\begin{eqnarray}
\big({f_{m,n}^\mp}\,,\,{f_{k,l}^\pm}\big)_{L^2} \= \int\,\dd x~\dd
t~f_{n,m}^\pm(t,x)\,f_{k,l}^\pm(t,x) \= 
\delta_{n,l}\,\delta_{m,k} \ .
\end{eqnarray}
The normalization relation (\ref{fpmnorm}) is computed in an analogous way.

\bigskip

\section*{Note added}

\noindent
After this paper was submitted for publication, the
preprint~\cite{Zahn10} appeared with some critiques of our approach
in~\cite{Fischer08}. In particular, a counterexample to our
Theorem~4.2 is found, casting doubt on our choice of domain $\Phi$. While this critique is fully justified, we have
amended our calculations, and found that both our
usage of and conclusions infered from the matrix basis are still in
fact valid. Briefly, one splits the Minkowski space action functional
$S_{\rm M}$ at $g=0$ into two parts as
\begin{eqnarray}
S_{\rm M}\= \mbox{$\frac{1}2$}\,\big(S^{(+\epsilon)}+S^{(-\epsilon)}\big) \
, \qquad S^{(\pm\,\epsilon)}\= S_{\rm M}\pm\ii\tan(\epsilon)\, S_{\rm E} \ ,
\end{eqnarray}
where $0<\epsilon<\frac\pi2$ and $S_{\rm E}$ is the euclidean
action at $g=0$. In the same manner that $S_{\rm E}$ and $S_{\rm M}$ can be
related to the harmonic and inverted harmonic oscillators,
respectively, the actions $S^{(\pm\,\epsilon)}$ are related to the
\emph{complex harmonic oscillator} with hamiltonian
$\frac12\,\big(p^2-\e^{\mp\, 2\ii\epsilon}\, q^2\big)$. In contrast to
the inverted harmonic oscillator, this family of hamiltonians have
discrete eigenvalues given by the harmonic oscillator spectrum scaled
by $\pm\ii\e^{\mp\ii\epsilon}$, while its eigenfunctions span
$L^2(\R)$ and have the usual star-product projection property. Path integral quantisation can now be easily carried out,
leading to the results summarised above in the limit $\epsilon\rightarrow0$.
The details of our modified analysis, together with various
applications to the computation of propagators in our model, will
appear in a forthcoming paper~\cite{Fischer10}. As stressed
in~\cite{Langmann02,Fischer08} and reviewed above, the existence of the matrix basis is the crux
of the proof of duality covariance at the \emph{quantum} level. It also led to the
original proof~\cite{Grosse04} of renormalizability of the
Grosse--Wulkenhaar model. From our point of view,
the matrix basis is taken as part of the definition of the
(regularized) duality covariant quantum field theory. We do not know
how to establish quantum duality using the basis of continuum
eigenfunctions, on which the analysis of~\cite{Zahn10} is (partly) based.

\bigskip

\section*{Acknowledgments}

\noindent
We thank A.P.~Balachandran, E.B.~Davies, K.~Fredenhagen,
H.~Grosse, H.~Steinacker and J.~Zahn for helpful comments, discussions
and correspondence. The work of RJS is supported in part by grant ST/G000514/1 ``String Theory
Scotland'' from the UK Science and Technology Facilities Council.

\bigskip


\end{document}